\documentclass[twocolumn]{IEEEtran}
\IEEEoverridecommandlockouts
\usepackage{amsmath,amssymb,amsfonts}
\usepackage{algorithmic}
\usepackage{graphicx}
\usepackage{textcomp}
\usepackage[OT1]{fontenc}
\usepackage{pifont}
\usepackage{caption}
\usepackage{array}
\usepackage{setspace}
\usepackage{makecell}
\usepackage{tabularx}
\usepackage{float}
\usepackage{tikz}
\usepackage[backend=biber, bibstyle=ieee, citestyle=numeric, date=year, sorting=nyt, doi=true]{biblatex}
\renewbibmacro*{volume+number+eid}{%
  \printfield{volume}%
  \iffieldundef{number}
    {}
    {\printfield{number}}%
  \setunit{\addcomma\space}%
  \printfield{eid}}

\DeclareFieldFormat[article]{volume}{#1}
\DeclareFieldFormat[article]{number}{\mkbibparens{#1}}
\usetikzlibrary{shapes.geometric, arrows}
\tikzstyle{startstop} = [rectangle, rounded corners, minimum width=2cm, minimum height=0.6cm, text centered, draw=black, fill=red!30]
\tikzstyle{process} = [rectangle, minimum width=2cm, minimum height=0.6cm, text centered, draw=black, fill=blue!30]
\tikzstyle{decision} = [diamond, minimum width=2.5cm, minimum height=0.6cm, text centered, draw=black, fill=green!30]
\tikzstyle{arrow} = [thick,->,>=stealth]
\usepackage{siunitx}
\usepackage{url}
\usepackage{multirow}
\usepackage{threeparttable}
\usepackage{booktabs}

\usepackage[dvipsnames,svgnames]{xcolor, colortbl}

\definecolor{LightCyan}{rgb}{0.88,1,1}
\def\BibTeX{{\rm B\kern-.05em{\sc i\kern-.025em b}\kern-.08em
    T\kern-.1667em\lower.7ex\hbox{E}\kern-.125emX}}
\let\svthefootnote\thefootnote
\newcommand\freefootnote[1]{%
  \let\thefootnote\relax%
  \footnotetext{#1}%
  \let\thefootnote\svthefootnote%
}
\begin{document}
\title{A Model Fusion Approach for Enhancing Credit Approval Decision Making
}

\author{
    \IEEEauthorblockN{
        Yuanhong Wu\IEEEauthorrefmark{1}\IEEEauthorrefmark{4}, 
        Jingyan Xu\IEEEauthorrefmark{1}\IEEEauthorrefmark{4},  
        Wei Ye\IEEEauthorrefmark{2}, 
        Christina Schweikert\IEEEauthorrefmark{3},
        and D. Frank Hsu\IEEEauthorrefmark{1},~\IEEEmembership{Senior Member, IEEE}\\
    }
    \IEEEauthorblockA{
        \IEEEauthorrefmark{1}Laboratory of Informatics and Data Mining, \\
        Department of Computer and Information Science, Fordham University, New York, NY 10023, USA\\
    }
    \IEEEauthorblockA{
        \IEEEauthorrefmark{2}Department of Economics, Fordham University, New York, NY 10023, USA\\
    }
    \IEEEauthorblockA{
    \IEEEauthorrefmark{3}CSMS Division, St. John’s University, Queens 11439, NY, USA\\
    }
     \IEEEauthorblockA{
        \IEEEauthorrefmark{4}These authors contributed equally
    }
}

\maketitle
\begin{abstract}

Credit default poses significant challenges to financial institutions and consumers, resulting in substantial financial losses and diminished trust. As such, credit default risk management has been a critical topic in the financial industry. In this paper, we present Combinatorial Fusion Analysis (CFA), a model fusion framework, that combines multiple machine learning algorithms to detect and predict credit card approval with high accuracy. We present the design methodology and implementation using five pre-trained models. The CFA results show an accuracy of 89.13\% which is better than conventional machine learning and ensemble methods.

\end{abstract}



\section{Introduction}

\freefootnote{\textcopyright 2025 IEEE. Personal use of this material is permitted. Permission from IEEE is required for all other uses.}

Credit card approval is a vital task in retail banking, directly affecting both profitability and risk exposure. As credit markets grow and application volumes rise, banks face increasing pressure to make faster and more accurate decisions. In response, machine learning algorithms have been used to evaluate applicant creditworthiness.


Traditional models such as logistic regression, decision trees, and $k$-Nearest Neighbors (KNN) have been widely used, showing effectiveness in credit risk prediction and fraud detection. More recently, ensemble methods have been adopted to boost accuracy by combining the strengths of multiple base models.


Combinatorial Fusion Analysis (CFA) is a new paradigm for computational leaning and modeling, X-informatics, and model fusion \cite{hsu2006combinatorial}. CFA provides techniques for model fusion, using both rank and score combinations. It also defines a quantitative measure for cognitive diversity between classifiers and leverages this diversity in model fusion. CFA has been used to improve performance in decision-making tasks in various domains including portfolio management, information retrieval, chemoinformatics, intrusion detection, among others \cite{Hurley}. This study presents the first empirical application of CFA to enhance credit approval prediction task. In this paper, we demonstrate that rank-based model fusion can outperform conventional machine learning and ensemble methods.


The remainder of this paper is organized as follows: Section I introduces the research background and motivation; Section II reviews related work; Section III outlines the methodology, including CFA framework, base models, and the settings for training models; Section IV presents the results; and Section V concludes the study with a discussion for future work.

\section{Related work}


Credit card risk management has been a key research area since the widespread adoption of credit cards, with credit application evaluation remaining a central focus. Machine learning has played a prominent role in this domain since the 1980s \cite{carter1987assessing}, and its use has grown with advances in computation and data availability.

Hand and Henley \cite{hand1997statistical} provided a review of statistical classification methods used in consumer credit scoring, highlighting their assumptions, strengths, and limitations. Roy and Vasa \cite{roy2025transforming} presented a PRISMA-based systematic review of AI and machine learning applications in credit risk assessment, spanning traditional statistical methods through advanced deep learning models. They found that ensemble models—especially boosted trees and random forests—consistently outperform others in accuracy, while also highlighting emerging challenges such as fairness, model interpretability, and the handling of imbalanced datasets. Schmitt \cite{schmitt2022deep} also compared the performance of ensemble models and deep learning approaches, identifying gradient boosting machines (GBM) and deep neural networks as two dominant models in credit risk management. He evaluated their performance on three structured credit-scoring datasets and found that GBM generally outperforms deep learning in terms of AUC and computational efficiency, although deep models remain competitive in certain cases.

Quan and Sun \cite{quan2024credit} introduced a hybrid credit scoring model based on a Factorization Machine (FM) model,
which integrated linear modeling with collaborative filtering to capture feature interactions. Their method effectively models sparse categorical and continuous features, yielding improved predictive accuracy on credit application datasets. Dahiya et al. \cite{dahiya2016impact} investigated the impact of bagging ensemble on a multilayer perceptron (MLP) classifier for credit evaluation using the Australian credit dataset. The authors conclude that the bagging ensemble significantly improves the classification accuracy of the MLP model. Emmanuel et al. \cite{emmanuel2024machine} developed a credit risk prediction engine using a stacked classifier that combines Random Forest, Gradient Boosting, and XGBoost as base models. Their stacked ensemble outperforms individual classifiers like ANN, decision tree, and KNN.

Some papers studied how to make credit risk models more explainable, shifting the focus from pure predictive performance to model transparency. For example, Misheva et al. \cite{misheva2021explainable} applied both LIME and SHAP to explain the predictions of machine learning models trained on Lending Club data. Their work demonstrates how post-hoc explanation techniques can enhance model interpretability and support trust in automated credit decision systems.

Across almost all the papers we reviewed in credit risk management, very few models use ranking-based approaches for credit risk prediction. In this paper, we apply a model fusion paradigm, Combinatorial Fusion Analysis (CFA), proposed by Hsu et al. \cite{hsu2006combinatorial}, that integrates models using both rank and score information. It leverages the rank-score characteristic (RSC) function and cognitive diversity measures \cite{hsu2019cognitive, hsu2010rank} to exploit inter-model diversity. CFA has shown effectiveness in a variety of diverse domains \cite{alfatemi2025identifying, comparing, Jiang, wu2025bitcoin}. This study presents the first such application, introducing a novel model fusion perspective to credit risk management.

\section{Methodology}

This section covers five base models, method of Combinatorial Fusion Analysis, and model training with dataset provided by UCI Machine Learning Repository.

\subsection{Base Models}
\subsubsection{KNN (Model A)}


The $k$-Nearest Neighbors (KNN) algorithm is a non-parametric method used for classification and regression \cite{knn}. It identifies the $k$ closest data points in the feature space and assigns the majority class or average value. KNN relies on distance metrics such as Euclidean distance and assumes that similar instances are close to each other. It does not require prior assumptions about the data distribution.

\subsubsection{LDA (Model B)}


Linear Discriminant Analysis (LDA) is a supervised method for classification and dimensionality reduction \cite{lda}. It projects data onto a lower-dimensional space that maximizes between-class variance while minimizing within-class variance. 

\subsubsection{AdaBoost (Model C)}


Adaptive Boosting (AdaBoost) is an ensemble algorithm that combines multiple learners of varying performance to form a strong classifier \cite{adaboost}. It sequentially trains learners on weighted data, increasing focus on previously misclassified samples. Final predictions are made through a weighted majority vote. 

\subsubsection{Random Forest (Model D)}


Random Forest builds an ensemble of decision trees using bootstrapped data and random feature selection \cite{rf}. This approach introduces diversity among trees, improving generalization and reducing overfitting. Predictions are aggregated by majority vote or averaging.

\subsubsection{CNN (Model E)}


Convolutional Neural Networks (CNNs) are deep learning models designed to process grid-structured data like images and sequences \cite{cnn}. Convolutional layers with shared weights are used to detect spatial patterns. Pooling and fully connected layers are utilized to reduce dimensions and perform classification or regression.

\subsection{Combinatorial Fusion Analysis}

A scoring system is represented by the following three components: a score function, a rank function, and a rank-score characteristic (RSC) function. In the context of this work, the multiple scoring systems (MSS) considered are: A ($k$-Nearest Neighbors), B (Linear Discriminant Analysis), C (AdaBoost), D (Random Forest), and E (Convolutional Neural Network). For each customer who applies for credit from a bank, a model provides a score value that represents the probability that either the customer should be approved or rejected due to credit risk. 

These scores are normalized to fall within the [0,1] range, resulting in five scoring systems labeled A through E. For a given set of $n$ credit  card applicants $D=\{d_1,..., d_n\}$, each scoring system includes a score function ($s_A, s_B, s_C, s_D, s_E$) and a derived rank function ($r_A, r_B, r_C, r_D, r_E$), which ranks the applicants in the descending order of trustworthiness. Given the score and rank functions, the RSC function $f_A$ for a scoring system $A$ is defined as:

\vspace{-1em}

\begin{equation}
f_A(i)=(s_A\circ r_A^{-1})(i)=s_A(r_A^{-1}(i))
\end{equation}\vspace{-0.1em}for $i=1,...,n$ \cite{hsu2006combinatorial, hsu2010rank, IEEEComp2024, Hurley}.

Cognitive Diversity (CD) quantifies the difference between a pair of scoring systems in terms of how they score the same dataset items, in this case credit card applicants \cite{hsu2019cognitive}. It is defined as the root mean square difference between the RSC functions of two systems, across all applicants:
\vspace{-1em}

\begin{equation}
    CD(A,B)=\sqrt{\frac{\sum^{n}_{i=1}((f_A(i)-f_B(i))^2)}{n-2}}
\end{equation}

\vspace{-0.3em}

Diversity strength $DS(A_j)$ of scoring system $A_j$ is computed for a scoring system, $A_j$, as the average CD between all pairs of scoring systems that include $A_j$ \cite{Jiang}, which is defined as $DS(A_i) = \frac{\sum_{j=1, j\neq i}^t CD(A_i, A_j)}{t-1}$ \cite{Jiang}. 

Within the CFA framework, MSS can be combined using either score combination or rank combination. The fusion can be unweighted (equal weight for all systems) or weighted, with weights reflecting diversity \cite{IEEEComp2024, Jiang}:

\vspace{-1em}

\[
w_j =
\begin{cases}
\vspace{1mm}
1 & \parbox[t]{5cm}{\raggedright for average combination (AC)} \\
\vspace{1.3mm}
DS_j & \parbox[t]{5cm}{\raggedright for weighted combination by diversity strength (WCDS)}
\end{cases}
\]

\vspace{-0.5em}

Let $A_1,..., A_t$ be scoring systems with corresponding weights $w_1,..., w_t$, the score combination is computed as \cite{IEEEComp2024, Jiang}:

\vspace{-1em}

\begin{equation}
    s_{SC}(d_i)=\frac{\sum^t_{j=1}w_j*s_{A_j}(d_i)}{\sum^t_{j=1}w_j}
\end{equation}. 

\vspace{-1em}

\noindent Similarly, the rank combination is calculated as \cite{IEEEComp2024, Jiang}:

\vspace{-1em}

\begin{equation}
    s_{RC}(d_i)=\frac{\sum^t_{j=1}(\frac{1}{w_j})*r_{A_j}(d_i)}{\sum^t_{j=1}(\frac{1}{w_j})}
\end{equation}

\vspace{-0.8em}

This weighting method allows the fusion process to prioritize systems with higher performance and higher diversity relating to other scoring systems simultaneously.

\subsection{Model Training}

This study uses the Australian Credit Approval dataset from the UCI Machine Learning Repository, which contains 690 instances with 14 attributes of various types. Each instance is labeled to indicate whether a credit application is approved (1) or denied (0). The mix of numerical, nominal, and binary features makes the dataset well-suited for evaluating models on heterogeneous inputs.

Each instance in the dataset represents an individual's credit profile. The attributes for each applicant include: 6 numerical attributes (e.g., applicant's age, credit amount) and 8 categorical attributes, of which: 3 are binary and 5 are nominal (multi-category). The target label is binary (approved or not approved). We confirmed that the dataset is relatively balanced, with approximately 55\% approvals and 45\% denials, reducing the risk of bias in model training.

In order to analyze features of different types, we apply the following encoding strategies. For numerical features, z-score normalization (zero mean, unit variance) is used. Standardization improves model performance by ensuring that features are on a comparable scale, which is particularly beneficial for distance-based models (e.g., KNN) and gradient-based optimizers used in neural networks. The nominal features are converted using one-hot encoding, which creates a set of binary variables for each category of the feature. Finally, we retain the binary features in their original format. All transformations are implemented using \texttt{scikit-learn} preprocessing tools to ensure reproducibility.

After preprocessing, the dataset is split into 80\% training and 20\% testing sets using stratified sampling to preserve label proportions. This allows for robust model evaluation without introducing distributional shifts between train and test subsets. Following the split, we train five base models as outlined in the previous section. Each model is optimized using randomized hyperparameter search with 5-fold cross-validation on the training set. The models output class probability scores rather than binary predictions. These probability scores capture each model’s confidence and are more informative than binary decisions. The probability scores will be used to construct scoring systems, which will then be combined using CFA fusion techniques.

\begin{figure}
    \centering
    \includegraphics[width=0.9\linewidth]{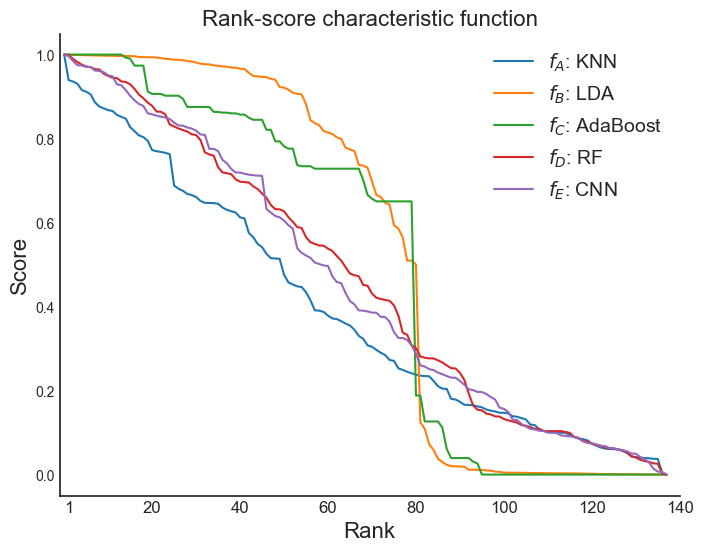}
    \caption{Rank-score functions for base models A, B, C, D, and E.}
    \label{fig:rsc_function}
\end{figure}

\begin{figure}
    \centering
    \includegraphics[width=\linewidth]{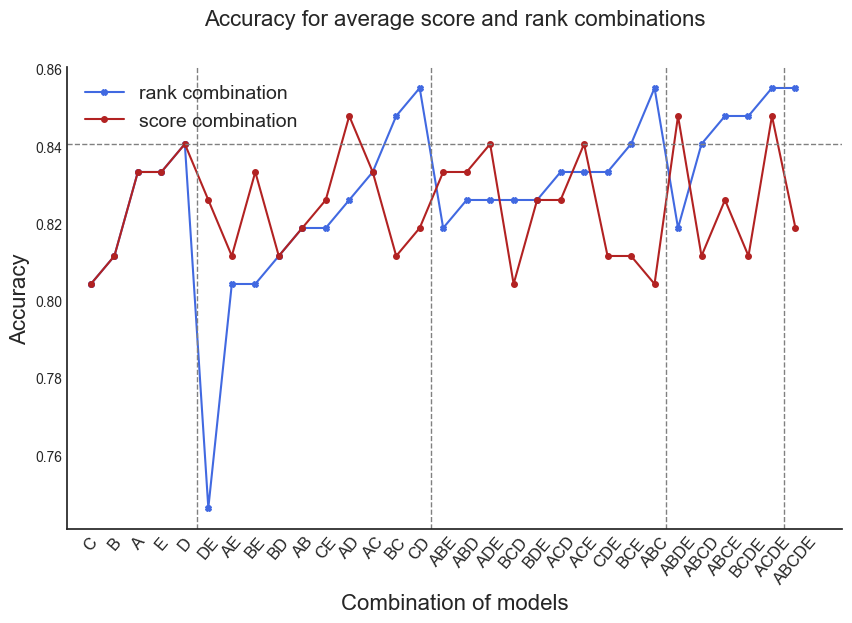}
    \caption{Accuracy for average score and rank combinations. 
    }
    \label{fig:ave_plot}
\end{figure}

\begin{figure}
    \centering
    \includegraphics[width=\linewidth]{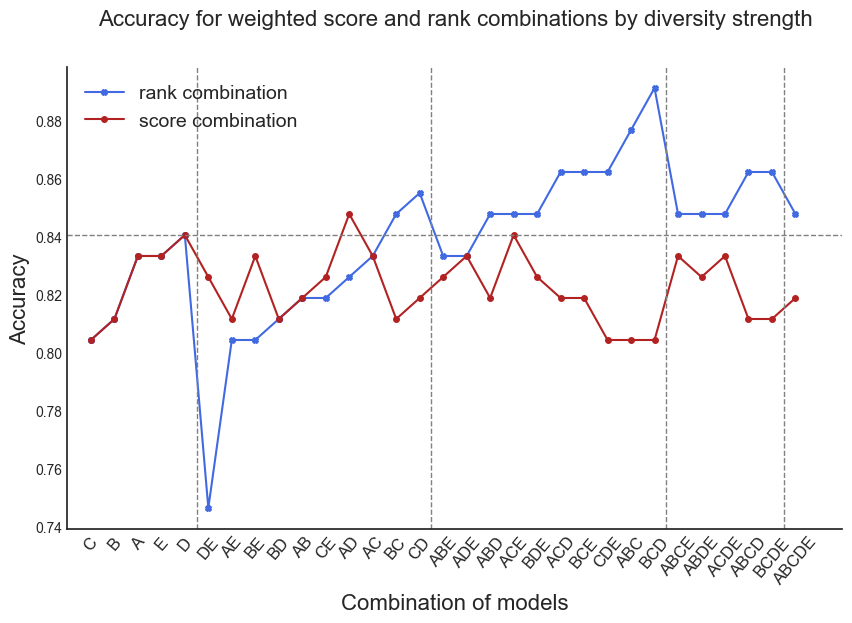}
    \caption{Accuracy for weighted score and rank combinations by diversity strength. 
    }
    \label{fig:ds_plot}
\end{figure}

\begin{table*}[t]
\centering
\small
\renewcommand{\arraystretch}{1.2}
\begin{tabularx}{0.934\linewidth}{|>{\centering\arraybackslash}m{1.7cm}|
                                >{\centering\arraybackslash}m{1.7cm}|
                                >{\centering\arraybackslash}m{2.3cm}|
                                >{\centering\arraybackslash}m{2.5cm}|
                                >{\centering\arraybackslash}m{3.2cm}|
                                >{\centering\arraybackslash}m{2.9cm}|
                                }
\hline
\multirow{2}{*}{} 
& \multirow{2}{*}[-0.3em]{\centering\makecell{Best\\Single Model}}
& \multicolumn{4}{c|}{CFA $k$-Combinations (k=2, 3, 4, 5)}  \\ \cline{3-6}
& & Best 2-Model Comb.
  & Best 3-Model Comb.
  & Best 4-Model Comb.
  & Best 5-Model Comb. \\ \hline
Score Comb.
& D (0.8406) 
& AD (AC, WCDS: \textbf{0.8478}) 
& ADE (AC: 0.8406), ACE (AC, WCDS: 0.8406) 
& ABDE (AC: \textbf{0.8478}), ACDE (AC: \textbf{0.8478}) 
& ABCDE (AC, WCDS: 0.8188)  \\ \hline
Rank Comb.  
& D (0.8406) 
& CD (AC, WCDS: \textbf{0.8551}) 
& BCD (WCDS: \textbf{0.8913}) 
& ABCD (WCDS: \textbf{0.8623}), BCDE (WCDS: \textbf{0.8623}) 
& ABCDE (AC: \textbf{0.8551})  \\ \hline
\end{tabularx}
\caption{Performance of $k$-combinations ($k$=2, 3, 4, 5). Numbers higher than the best single model performance are bold.}
\label{tab:ac}
\end{table*}
In this work, we adopt accuracy as the performance metric. This choice is motivated by the operational priorities of credit card centers, which are primarily concerned with making correct approval or denial decisions for credit card applications. Accuracy directly reflects the quality of this decision-making process. Moreover, accuracy is widely used in related studies, allowing for straightforward comparison.

As previously described, the Combinatorial Fusion Analysis (CFA) framework leverages both score-based and rank-based combinations. The score refers to the probabilities, which are typically converted to predictions using a threshold of 0.5. In contrast, rank-based fusion transforms probabilities into ordinal rankings, where lower-ranked applications are considered more reliable. A lower rank implies lower credit risk and a higher likelihood of approval.

To create binary labels from the ranks, we use the number of positive instances (i.e., approved applications) in the ground truth as a threshold. For example, if the dataset contains 100 approved cases, we label the top 100 ranked applications as class 1 (approved), and the remaining as class 0 (not approved). This approach is intuitive and practical, as it assumes that the lowest-ranked applicants are those most likely to be creditworthy, aligning well with real-world risk assessment practices.

\section{Results}

Combinatorial Fusion Analysis is a model fusion framework that provides an efficient and robust approach to combine multiple scoring systems. It integrates the strengths of base models by exploring combinations at both the score and rank levels, as well as through weighted combinations. By taking advantage of the diversity of scoring and ranking behavior, CFA significantly enhances the potential of model fusion. According to theoretical findings, rank-based combinations can outperform score-based combinations under certain conditions, which provides additional motivation to consider both representations \cite{comparing}.

In this study, we employ five base models and systematically examine their combinations. For each combination size $t \in \{2, 3, 4, 5\}$, we consider all $\binom{5}{t}$ possible combinations. For example, there are 10 distinct 2-model combinations ($\binom{5}{2} = 10$). In total, this results in $2^5 - 1 - 5 = 26$ unique model combinations, excluding individual models and the empty set. Each combination is evaluated using both score-based and rank-based fusion methods, and further combined under two different weighting schemes: unweighted average and diversity-based weighted average. This leads to a total of $26 \times 2 \times 2 = 104$ model fusion cases.

To compare across different combination strategies, we present the performance of model combinations in two separate figures. Figures \ref{fig:ave_plot} and \ref{fig:ds_plot} correspond to the results from averaging and diversity-weighted fusion, respectively. Before analyzing the performance of these scoring system combinations, we first examine the rank-score characteristic (RSC) functions of the five base models, as shown in Figure \ref{fig:rsc_function}. The RSC functions provide a clear illustration of model diversity, which plays a crucial role in enhancing model fusion performance.

In Figure \ref{fig:rsc_function}, the x-axis represents the rank of instances in the test dataset, while the y-axis shows the normalized scores, which are the raw predicted probabilities scaled to the range $[0, 1]$. These normalized scores are sorted in descending order, so that rank 1 corresponds to the highest score.



From Figure \ref{fig:rsc_function}, we observe that the five base models exhibit noticeable diversity in their rank-score profiles. Most models show clear separation, with the exception of Models D and E, which appear closely aligned. In contrast, models that differ more in their rank-score behavior offer greater diversity, an important factor in boosting the effectiveness of model fusion.

Figure \ref{fig:ave_plot} compares the classification accuracy of average score and rank combinations across all model groups. The x-axis displays the five individual base models along with all 26 possible combinations of these models. Among the individual models, Model D (Random Forest) achieves the highest accuracy of 0.84, which is indicated by a horizontal dashed line serving as a reference threshold. Any model combination exceeding this line demonstrates an improvement over the best-performing individual model.

In the group of 2-model combinations, rank-based fusion yields two combinations (BC and CD) that outperform the best individual model, whereas score-based fusion identifies only one such combination, AD. Notably, the rank combination CD exhibits a greater performance gain than the score combination AD. This can be attributed to the inclusion of Model D (the most accurate model) and its relatively large diversity compared to Model C, which enhances the effectiveness of the ensemble.

For 3-model combinations, only one rank-based fusion exceeds the threshold, while the best score-based combination merely matches the performance of the top individual model. In the 4-model combination group, rank-based fusion produces three combinations with improved accuracy, compared to two from the score-based method. Finally, when all five base models are combined, rank-based fusion again surpasses the individual benchmark, whereas score-based fusion does not.

\begin{table*}[t]
\centering
\small
\begin{tabular}{p{3.3cm}|p{10.5cm}|p{1.5cm}}
\hline \hline
Category & Method & Accuracy \\
\hline
\multirow{3}{3.5cm}{Hybrid or Ensemble approach} & MLP-based bagging ensemble \cite{dahiya2016impact} & 0.8875 \\[1ex]

                            & Factorization Machine \cite{quan2024credit}& 0.8844 \\ [1ex]
                            & Stacking RF, Gradient Boosting, and XGB \cite{emmanuel2024machine}& 0.8623 \\
                            
\hline
\multirow{6}{3.5cm}{CFA} & Average score combination           (AD, ABDE, ACDE) & 0.8478 \\[1ex]
                & Average rank combination (ABC)  & 0.8551 \\[1ex]
                & Weighted score combination by DS (AD) & 0.8478\\[1ex]
                & Weighted rank combination by DS (BCD) & \textbf{0.8913} \\
\hline \hline
\end{tabular}
\caption{Comparing CFA with other single model and ensemble models.}
\label{tab:leaderboard}
\end{table*}
Overall, rank-based fusion demonstrates superior performance compared to score-based fusion, suggesting that rank averaging provides a better strategy for model combination in this setting.

Figure \ref{fig:ds_plot} presents the classification accuracy of weighted score and rank combinations, where the weights are derived from model diversity. For 2-model combinations, the results are identical to those from the average-based approach in Figure \ref{fig:ave_plot}: rank-based fusion identifies two combinations (BC and CD) that surpass the best individual model, while score-based fusion yields only one (AD). However, the advantage of rank fusion becomes more apparent as we move to 3-, 4-, and 5-model combinations. In these groups, the accuracy curves for rank-based combinations consistently lie above those of score-based combinations. Notably, score fusion fails to produce any combinations that outperform the best individual model, while most rank combinations show clear improvement. Several rank-based combinations, such as ABC and BCD, even exceed an accuracy of 0.86.


Table \ref{tab:ac} compares the performance of the best individual model with that of the best model combinations of varying fusion sizes obtained using the CFA method. Each cell corresponding to the CFA results contains the following information: (a) outside the parentheses indicates the models combined; (b) the first entry inside the parenthese indicates types of combination, for example, "AC" represents average combination, while "WCDS" denotes weighted combination by diversity strength; (c) the second entry in the parenthese shows the accuracy of the combination. Bolded accuracies indicate performance improvements over the best individual model. Overall, we observe that CFA-based model combinations outperform individual models across both score-based and rank-based fusion methods, as well as across different combination sizes. Additionally, rank fusion improves accuracy across all four combination sizes, whereas score fusion shows better performance in only two of them.

Among all combinations, the most accurate model set is BCD from weighted rank combination by diversity strength, consisting of LDA, AdaBoost, and Random Forest. This combination achieves an accuracy of \textbf{0.8913}, representing a substantial gain over the best individual model (Random Forest with 0.84). This nearly $5\%$ improvement is particularly significant given the high baseline accuracy, where even a $1\%$ gain is typically challenging to achieve. As shown in Figure \ref{fig:rsc_function}, the DS (diversity strength) weighted combination of models B, C, and D achieved the highest performance in part due to their diversity.

Previously, we compared the performance of the CFA-based model combinations with their individual base models. We now extend this analysis by evaluating the CFA results against several established hybrid and ensemble approaches. The comparison is summarized in Table \ref{tab:leaderboard}.

All the accuracy of our base models are lower than that of the hybrid or ensemble methods listed in Table \ref{tab:leaderboard}. In the table, Dahiya et al. \cite{dahiya2016impact} employed a bagging ensemble of multilayer perceptrons (MLP) and achieved an accuracy of 0.8875. Quan et al. \cite{quan2024credit} proposed a hybrid Factorization Machine model, which integrated linear modeling with collaborative filtering to capture feature interactions. This model achieved an accuracy of 0.8844 on the same dataset. Additionally, Emmanuel et al. \cite{emmanuel2024machine} introduced a stacking ensemble comprising Random Forest (RF), Gradient Boosting, and XGBoost (XGB), yielding an accuracy of 0.8623.

Turning to the results from the CFA framework, we evaluated four different fusion strategies and reported the highest accuracy achieved under each combination technique, along with the corresponding set of fused models. Many of the CFA combinations achieved accuracies around 0.85. The highest observed accuracy was 0.8913, obtained from the weighted rank combination based on diversity strength, using models B, C, and D. This performance exceeds the hybrid and ensemble models included in Table \ref{tab:leaderboard}.

These results underscore the effectiveness and flexibility of the CFA framework. By systematically exploring and leveraging the complementary strengths of diverse models, CFA not only matches but often outperforms traditional ensemble and hybrid approaches. Its model fusion strategy allows tailored combinations that adapt to the specific diversity or performance patterns of base classifiers, which makes CFA a powerful tool for enhancing predictive accuracy in complex decision-making tasks such as credit card approval prediction.

The bias, transparency, and fairness issues arising from a single model for task like this may be too critical given the regulatory and societal implications of credit approval. The use of rankings makes it easier to explain to the potential applicants why their applications got approved or rejected. The cognitive diversity and aggregation techniques can also detect and effectively diminish biases. And these help position CFA within the growing movement toward responsible AI in finance.

\section{Conclusion and Future work}

In this study, we applied the Combinatorial Fusion Analysis (CFA) framework to enhance credit card approval prediction. We selected five diverse base models: KNN, LDA, AdaBoost, Random Forest, and CNN, as inputs to the CFA framework. These models were fused using two weighting schemes: average combination and diversity-based weighted combination.

Our experimental results demonstrate that CFA-based model fusions significantly outperform individual base models. CFA configurations achieved higher accuracy than state-of-the-art hybrid and ensemble methods reported in recent literature. Notably, the weighted rank combination by diversity strength using models BCD (LDA, AdaBoost, RF) reached an accuracy of 0.8913, surpassing the existing benchmarks. 

A key strength of CFA lies in its use of not only scores but also ranks. Our results suggest that rank-based combinations, as implemented in the CFA framework, can lead to unexpectedly strong improvements in predictive performance, especially given diverse models/algorithms. This rank-oriented approach and quantitative measure for model diversity captures complementary information that may not be fully reflected in score-based fusion. Therefore, we advocate for the broader adoption of rank-based fusion strategies, and highlight CFA as a practical and flexible framework for achieving this, especially in applications like credit risk prediction, where decision quality is critical.

Looking ahead, we identify two future research directions. First, we plan to explore a multi-layer CFA architecture, where the output of one layer of model fusion is used as input to a second-layer fusion. In this way, the initial 104 combinations could be reduced to five higher performing ensembles by accuracy and diversity criteria, which are then recombined to yield potentially even better results. Preliminary experiments suggest that this multi-layer design can lead to further accuracy improvements \cite{Hurley}.

Second, we aim to integrate online learning capabilities into the CFA framework. This would allow banks to dynamically update their models as new application data arrives, eliminating the need to retrain models from scratch.

On the other hand, accuracy alone can't reflect asymmetric error costs or class imbalance. Therefore, the future work must extend to metrics that can capture these traits. To validate robustness and generalizability, we should also incorporate additional datasets that's different from our current one in terms of imbalance distribution, structure, and preferably some other domain field that isn't credit card approval but credit risk related.


\printbibliography
\end{document}